\documentclass[aps,preprint,floatfix,amssymb,amsmath,tighten,prb,endfloats]{revtex4}

\usepackage[dvips]{graphicx}
\graphicspath{{.}}
\newcommand{\tf}{\ensuremath{T_\mathrm{f}}}

\newcommand{\rmf}{\ensuremath{\mathrm{f}}}

\newcommand{\beq}{\begin{equation}}
\newcommand{\eeq}{\end{equation}}

\makeatletter
\renewcommand\@biblabel[1]{#1.}
\makeatother

\makeatletter
\def\@cite#1#2{$(#1\if@tempswa , #2\fi)$}
\makeatother

\begin{document}


\title{\mbox{Capillarity-like growth of protein folding nuclei}}

\author{Xianghong Qi and John J. Portman}
\affiliation{
Department of Physics, Kent State University, Kent, OH 44242
}
\date{December 13, 2007}


\begin{abstract}
A full structural description of transition state ensembles
in protein folding includes the specificity of the ordered residues
composing the folding nucleus as well as spatial density.  To our
knowledge, the spatial properties of the folding nucleus and interface of
specific nuclei has yet to receive significant attention. We analyzed
folding routes predicted by a variational model in terms of a
generalized formalism of the capillarity scaling theory that assumes
the volume of the folded core of the nucleus grows with chainlength as
$V_\rmf \sim N^{3\nu}$. For 28 two-state proteins studied, the scaling
exponent $\nu$ ranges from 0.2 to 0.45 with an average of $0.33$. 
This average value corresponds to packing of rigid objects,
though generally the effective monomer size in the folded core is
larger than the corresponding volume per particle in the native state
ensemble.  That is, on average the folded core of the nucleus is found
to be relatively diffuse.  We also studied the growth of the folding
nucleus and interface along the folding route in terms of the density
or packing fraction. The evolution of the folded core and
interface regions can be classified into three patterns of growth
depending on how the growth of the folded core is balanced by changes
in density of the interface. Finally, we quantify the diffuse versus
polarized structure of the critical nucleus through direct calculation
of the packing fraction of the folded core and interface regions. Our
results support the general picture of describing protein folding as
the capillarity-like growth of folding nuclei.
\end{abstract}
\maketitle






The modern theory of protein folding describes the mechanism
for folding as an entropic bottleneck arising from the decreasing
number of accessible pathways available to a protein as it becomes
ordered.~\cite{leopold:onuchic:92,bryngelson:wolynes:95} The
collection of partially ordered conformations corresponding to this
bottleneck region is known as the transition state ensemble or
critical folding nucleus.~\cite{onuchic:socci:95} Although it is
common to focus on the degree of native-like order of specific
residues, a complete description of the protein folding mechanism also
includes the spatial properties such as size or density of the
transition state ensemble.  Indeed, shortly after characterizing the
transition state ensemble of CI2, Fersht summarized the structure of
the critical nucleus by a spatial description through the proposal of
the \textit{nucleation-condensation}
mechanism.~\cite{itzhaki:fersht:95} This critical nucleus can be
thought of as an expanded, partially ordered version of the native
state ensemble with concomitant longranged tertiary and local
secondary structure.  It is now clear that while diffuse nuclei appear
to be the general rule, some nuclei are less diffuse than
others.~\cite{alm:baker:02} Polarized nuclei have highly structured
residues which are spatially clustered in the native structure, while
the rest of the residues show little definite order.
~\cite{martinez:serrano:99,riddle:baker:99,jager:gruebele:01,garcia-mira:schmid:04}
Such nuclei are similar to the capillarity approximation in
homogeneous nucleation in which the free energy of a stable phase
droplet is separated from the metastable phase by a very sharp
interface.\cite{finkelstein:badretdinov:97,wolynes:97} Exploiting this
analogy, Wolynes describes a nucleus with capillarity-like order in
which the interface surrounding a relatively folded core is broadened
by wetting of partially ordered residues.~\cite{wolynes:97} In this
picture, folding can be described as the growth of the folding
nucleus: a wave of order moving across the protein as the edge of the
nucleus expands to ultimately consume the entire
molecule.\cite{wolynes:97,oliveberg:wolynes:05}

\begin{figure}[p]
\centering
\includegraphics[width=2.50in]{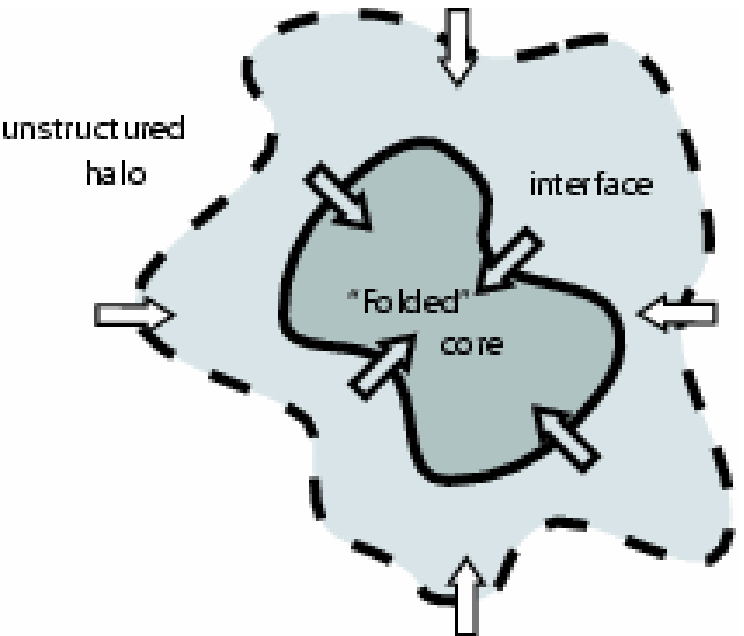}
\caption{Illustration of folding nucleus: folded core, interfacial region, and unfolded halo.
Growth of the nucleus can be characterized by fluxes entering the folded core and interfacial regions.
\label{fig:nucleus}}
\end{figure}

The extended partially ordered interface of a capillarity-like ordered
nucleus separates space into three regions: a folded core, a partially ordered interface
region, and unfolded halo (see Fig.~\ref{fig:nucleus}).  In this
paper, we monitor the structural development of the nucleus along the
folding route through the evolution of the packing fraction of the
folded core and the interface.  As shown in
Fig.~\ref{fig:nucleus}, growth of the nucleus can be described by
fluxes of residues passing through two moving surfaces: one surface
separates the folded core and interface, and the other surface
separates the interface region and the unfolded halo. As the protein
folds, the evolution of the interface is determined by the interfacial
volume and the net flux of residues entering the interface.


Our analysis is based on folding routes calculated for 28 two-state
proteins from a cooperative variational model described in
\citealp[Ref.][]{qi:portman:07}.  We note this model includes neutral
cooperativity due to repulsive excluded volume interactions.  This
form of cooperativity has been shown to broaden the range of barrier
heights allowing direct comparison between calculated and measured
folding rates.\cite{qi:portman:07} Not surprisingly, cooperativity
tends to sharpen the interface between folded and unfolded
regions. Nevertheless, the interface from this model is generally not
nearly as sharp as a strict capillarity description in which a residue
can be clearly identified as being either completely folded or
completely unfolded as some other analytic models
assume.~\cite{galzitskaya:finkelstein:99,alm:baker:99,munoz:eaton:99}
In fact, an unbiased analysis of the spatial properties of the folding
nucleus fundamentally depends the model's ability to describe partial
order.

\section*{Capillarity-like growth of folding nucleus}
\textbf{Capillarity picture of folding nuclei.}~%
The capillarity approximation of folding nuclei is based on classical
nucleation theory of first order phase transition
kinetics.~\cite{bryngelson:wolynes:90,wolynes:97}.  
Within the capillarity approximation, 
the free energy of a nucleus with volume $V_\rmf$ and
surface area $A_\rmf$ can be written as a sum of two terms
\beq\label{eq:fcap} 
F = -\Delta f V_\rmf + \gamma A_\rmf,
\eeq
where $\Delta f$ denotes the bulk free energy difference per unit
volume between the unfolded and folded ensembles, and $\gamma$
is the surface tension between the folded and unfolded regions. 

A folded core with native-like density has a volume per monomer
independent of its size.  Relaxing this assumption, we assume
that the number of residues in the folded
core, $N_\rmf$, scales with and its volume, $V_\rmf$, according to
\beq\label{eq:vf}
V_\mathrm{f} = b^3 N_\mathrm{f}^{3\nu}.
\eeq
Here, $\nu$ is the scaling exponent associated with the lengthscale of
the folded core $R \sim b_0 N_\rmf^{\nu}$, and $b^3$ is a geometry
dependent elementary volume proportional to the monomer volume,
$b_0^3$.  The free energy of a folded nucleus with $N_\mathrm{f}$
residues then has the form:~\cite{wolynes:97}
\beq\label{eq:free_energy} 
F(N_\mathrm{f}) = -\Delta f b^3
N_\mathrm{f}^{3\nu}+ \gamma b^2 N_\mathrm{f}^{2\nu}, 
\eeq 
At the folding transition temperature, $\tf$, finite size depression
of the surface energy suggests that $\gamma \sim \Delta f b N^{\nu}$
where $N$ is the number of monomers in the protein.  The maximum of
the free energy occurs at $N_\mathrm{f}^{\dagger}=(2/3)^{1/\nu}N$,
giving the size of the critical nucleus, and the associated free
energy barrier scales as $\Delta F^{\dagger}\sim N^{2\nu}$.  If we
assume that the folded core has native like packing, $\nu= 1/3$ and
$b^3$ is the native-like volume per monomer, so that
$N_\mathrm{f}^{\dagger}=(2/3)^{3}N$ and $\Delta F^{\dagger}\sim
N^{2/3}$.~\cite{wolynes:97,finkelstein:badretdinov:97}

Simulations and alternative theoretical considerations also suggest
that barrier height (logarithm of the folding time) scales sublinearly
on chainlength, $\Delta F^{\dagger} \sim
N^p$ with $0\le p \le 1$.~\cite{thirumalai:95,koga:takada:01,finkelstein:badretdinov:97,wolynes:97}
Direct analysis of folding rate data to determine the scaling exponent $p$
encounters the difficulty that the range of $N$ is too small to
distinguish between different values of $p$.%
\cite{ivankov:finkelstein:03,galzitskaya:finkelstein:03,ivankov:finkelstein:04,li:thirumalai:04,naganathan:munoz:05}.
So while it may be reasonable to expect that the scaling of the barrier
height with chainlength is universal for sufficiently large proteins,
the size of a typical two state proteins ($\sim$ 100 amino acids) may
well be too small to be governed by this generic behavior. In this
case, both specificity and size of these smaller proteins should
generally determine the properties of the critical nuclei. In this
paper, we assume that Eq.~\ref{eq:vf} is valid to describe the growth
of the nucleus in all the two state proteins, but the exponent $\nu$
and volume $b^3$ are allowed to be protein specific.\\

\noindent\textbf{Characterizing the folded core and the interface.}~%
In the variational model considered in this paper, partially ordered
configurations are described by a variational Hamiltonian,
$\mathcal{H}_0$, corresponding to a stiff polymer chain inhomogeneously
constrained to the native structure.  Since this model is described in
detail in Ref.~\cite{qi:portman:07}, we focus here on how to define folded
core, interface, and unfolded regions along the calculated folding
route. This is not as straight-forward as one might expect because the
concept directly couples specificity of the nucleus with
the spatial density.

We characterize the degree of structure of each residue by the extent
of localization about the native structure
$\{\mathbf{r}^{\mathrm{N}}\}$, 
$\rho_i = \langle \exp(-\alpha^{\mathrm{N}}(\mathbf{r}_i -\mathbf{r}_i^{\mathrm{N}})^2\rangle_0$, with $\alpha^{\mathrm{N}}=0.1$. Here, the subscript denotes the average
with respect to the Boltzmann weight with $\mathcal{H}_0$.  Denoting
the native density at the globule and native state by
$\rho_i(\mathrm{G})$ and $\rho_i(\mathrm{N})$, respectively, we
consider the normalized density
\beq\label{eq:native_density} 
\tilde{\rho}_i =\frac{\rho_i -\rho_i(\mathrm{G})}{\rho_i(\mathrm{N}) -\rho_i(\mathrm{G})}
\eeq 
as a set of order parameters characterizing the folding of each
residue.  Progress along the folding route can be monitored by the
global structural parameter $Q= 1/N\sum \tilde{\rho}_i$.

\begin{figure}[p]
\centering
\includegraphics[width=2.50in]{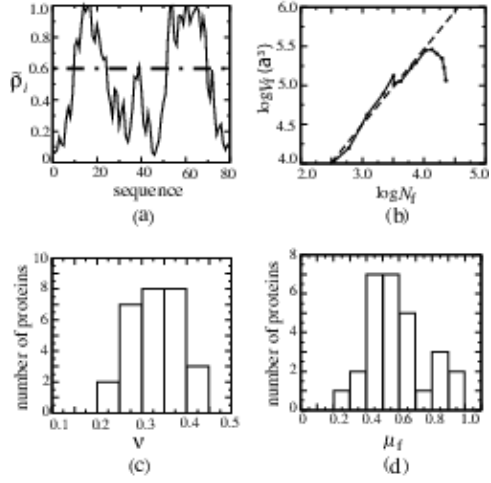}
\caption{Scaling of the folded core with number of monomers. (a-b) correspond
to $\lambda-$repressor (1lmb). (a) Residues
with native density $\tilde{\rho_i}>0.6$ (indicated by the dashed line)
define a fiducial set of folded residues.
(b) Linear fit of $\log V_\rmf$ vs. $\log N_\rmf$ (dashed line) gives 
the exponent of $V_f \sim N_f^{3\nu}$. In this example, 
the fitting equation is $y=5.6 + 0.97x$, so that $\nu=0.32$ and $b^3 =5.6a^3$,
$a=3.8\AA$ is the average distance between the $\alpha$ carbons. (c) Histogram
of scaling exponent $\nu$ for 28 proteins; (d) Histogram of the packing
fraction of the folded core of the critical nucleus at $\tf$.
\label{fig:dist_exponent}}
\end{figure}

The normalized native densities are used to define a
fiducial set of folded residues, $\{F\}$, with $\tilde{\rho}_i > 0.6$, as shown in 
Fig.~\ref{fig:dist_exponent}a.
Next, we define the spatial region of the folded core through the
relative contribution of the density of the folded residues in $\{F\}$,
$n_\mathrm{f}(\mathbf{r}) = \sum_{\{\mathrm{F}\}}
\langle \delta (\mathbf{r} - \mathbf{r}_i)\rangle_0$,
to the total density, $n(\mathbf{r}) = \sum_{i=1}^N
\langle \delta (\mathbf{r} - \mathbf{r}_i)\rangle_0$. The spatial
extent of the folded core and interfacial regions in this analysis is
determined by an indicator function
\beq\label{eq:rel_density}
\tilde{n}(\mathbf{r}) = \frac{n_{\mathrm{f}}(\mathbf{r})}{n(\mathbf{r})}.
\eeq
which ranges from  $0 \le \tilde{n}(\mathbf{r}) \le 1$.
We define the folded core region,
$\mathcal{V}_\mathrm{f}$,
as the points $\{\mathbf{r}_\rmf\}$ for which the density of the fiducial folded residues
contributes at least 50\% to the total density
($\tilde{n}(\mathbf{r}) \ge 0.5$).
The number of residues in this folded
core region can be found by numerically integrating the density over
the core region, $N_\mathrm{f}
=\int_{\mathcal{V}_{\mathrm{f}}}n(\mathbf{r})\mathrm{d}\mathbf{r}$. The
volume of the core region is given by $V_\mathrm{f}
=\int_{\mathcal{V}_{\mathrm{f}}}\mathrm{d}\mathbf{r}$.

Similarly, the interfacial region, $\mathcal{V}_{\mathrm{int}}$, is
defined as the points $\{\mathbf{r}_\mathrm{int}\}$ for which $0.1 \le
\tilde{n}(\mathbf{r}_\mathrm{int}) < 0.5$.  The number of
interfacial residues and volume of the interface is given by
$N_\mathrm{int}
=\int_{\mathcal{V}_\mathrm{int}}n(\mathbf{r})\mathrm{d}\mathbf{r}$,
and $V_\mathrm{int}
=\int_{\mathcal{V}_\mathrm{int}}\mathrm{d}\mathbf{r}$, respectively.

The number of residues and the volume can be used to define a
mean packing fraction of the folded core and partially ordered
interface by
\beq\label{eq:packing_fraction} 
\mu_\mathrm{f} = \frac{N_\mathrm{f}}{ V_\mathrm{f}}v_0 \qquad \mbox{and}\qquad
\mu_\mathrm{int} = \frac{N_\mathrm{int}}{V_\mathrm{int}}v_0,
\eeq 
respectively.  Here, $v_0$ is the calculated volume of per particle of
the native structure at the folding transition temperature,
$T_{\mathrm{f}}$.  The growth of the nucleus can
be characterized by the way the packing fractions $\mu_\mathrm{f}$
and $\mu_\mathrm{i}$ change along the folding route.\\

\noindent\textbf{Growth of folding nucleus along the folding route.}~%
As illustrated in Fig.\ref{fig:dist_exponent}(a-b), the changes in
$N_\rmf$ and $V_\rmf$ along a folding route can be fit to
Eq.~\ref{eq:vf} to give an estimate of the scaling exponent $\nu$ for
each protein.  Fig.~\ref{fig:dist_exponent}(c) shows the distribution
of predicted $\nu$ from the folding routes obtained from the
variational model for 28 two-state proteins discussed in
~\citealp[Ref.][]{qi:portman:07}.  The predicted scaling exponent $\nu$
ranges between $0.2 \sim 0.42 $ with an average of $\nu =0.33 $.  The
mean exponent is very close to the the scaling associated with
close-packed rigid objects, $\nu = 1/3$.  For comparison, recent
detailed statistical models indicate that the scaling exponent for the
unfolded state of a protein is about ~\cite{jha:sosnick:05}
$\nu = 0.59$, while the folded state of a wide variety of proteins
suggest that proteins with less than 300 amino acids have compact
folded structures ($\nu = 0.3$), while larger proteins are less dense
($\nu = 0.4$).~\cite{xu:leitner:03}


The mean packing fraction of the core scales with the number of monomers 
as: 
\beq\label{eq:muf_scale}
\mu_\rmf = \frac{v_0}{b^3}N_\rmf^{1-3\nu}.  
\eeq 
For the close packing value $\nu = 1/3$,  $\mu_\rmf$ is independent of the 
number of monomers and the core retains native like density as it grows.
When $\nu > 1/3$, the core becomes less compact as
monomers are added to the core. This is the familiar
scaling from loosely packed or fractal objects. When $\nu < 1/3$, the
core density increases as more monomers are incorporated into the
core. This can be understood as the consolidation of structure in the
folded core as folding progresses.

Although the spatial structure of the critical folding
nucleus (transition state ensemble) is discussed in more detail later, it is
instructive to consider the value of the mean packing fraction of the
core here. Fig.~\ref{fig:dist_exponent}d shows the distribution of
packing fractions of the folded core evaluated at the maximum free
energy barrier between folded and unfolded states at $\tf$.  The
packing fraction has a wide range from 0.2 to 1.0.  While some of the
transition state nuclei have compact cores, the average
packing fraction is only 0.59. This means that although the growth of
a typical folded core corresponds to rigidly packed objects, a typical
transition state ensemble has a folded core with twice the volume as
the volume of same number of monomers in the native state
conformation ($b^3 \approx 2v_0$).
That is, the monomers composing the nucleus are typically much 
less localized than in the native state.

Fig.~\ref{fig:1lmb} illustrates a typical example of the 
growth of the folded core and the interface region.
Early in the folding, we see a small compact nucleus surrounded by a
partially folded interface. This small nucleus is partially ordered,
occupying about twice the volume as the corresponding residues in the
native state.  Structural fluctuations giving nuclei corresponding to
$Q < Q^{\dagger}$ are unstable with respect to the unfolded state due
to relatively large surface free energy cost associated with small nuclei, whereas
structural fluctuations with $Q > Q^{\dagger}$ will tend to evolve to
the folded state.  As the nucleus grows, the volume of the
nucleus evolves as interfacial regions are incorporated into the
core, while unfolded regions become part of the partially ordered
interface.\\

\begin{figure}[p]
\centering
\includegraphics[width=2.50in]{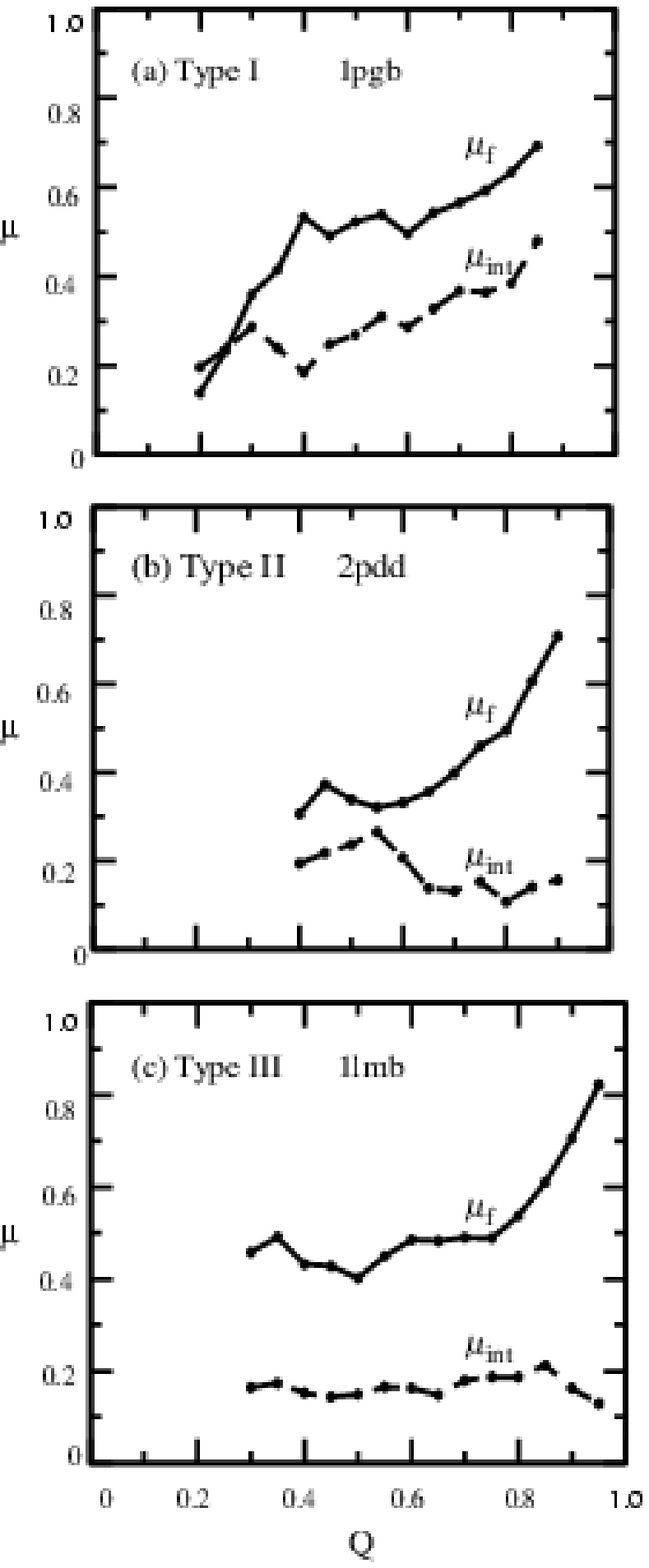}
\caption{Examples of three modes of growth of the nucleus.
Pattern A--C corresponds to (a--c), respectively.
Solid line corresponds to the mean packing fraction of the folded core, $\mu_\rmf$,
while
the dashed line corresponds to the mean packing fraction of the interface,
$\mu_\mathrm{int}$.
\label{fig:packing_group}}
\end{figure}

\noindent\textbf{Growth patterns of the nucleus:}~%
The structural growth of the folding nucleus can be understood as the
competition between growth of the folded core and the evolution of the
interface.  The flux of residues entering core through the interface
region controls the growth of the core, while the net flux of
residues entering interface region from the unfolded halo controls
the growth of the interface (see Fig.~\ref{fig:nucleus}).  We
characterize the evolution of the structure of the folding by the
changes in the packing fraction of the core and interface regions as a
function of $Q$. That is, we consider the signs of
\beq 
\dot{\mu}_\rmf (Q) =
\frac{\mathrm{d}\mu_\rmf}{\mathrm{d}Q} \quad \mbox{and} \quad
\dot{\mu}_\mathrm{int} (Q) =
\frac{\mathrm{d}\mu_\mathrm{int}}{\mathrm{d}Q}  
\eeq 
to identify different modes of growth.  From the two state proteins
used in this study, we can identify three distinct scenarios as
illustrated in Fig.~\ref{fig:packing_group}. 


\begin{itemize}
\item
Pattern A
\textit{(consolidation of core and interface).}
As shown
in Fig.~\ref{fig:packing_group}a, the density of both the core and
interface increase along the folding route ($\dot{\mu}_\mathrm{int} >
0$ and $\dot{\mu}_\mathrm{int} > 0$).  The size of the core increases
with $Q$, but $V_\rmf$ increases more slowly than $N_\rmf$ (see
Eq.~\ref{eq:muf_scale} with $\nu < 1/3$).  Similarly, $N_\mathrm{int}$
and $V_\mathrm{int}$ both increase with with $Q$ throughout much of the
growth. At larger $Q$, $V_\mathrm{int}$ reaches a maximum and
subsequently decreases rapidly as interfacial residues are consumed by
the folded core.

\item
Pattern B
\textit{(core consolidation dominated growth).}  As shown in
Fig.~\ref{fig:packing_group}b, the growth of the core is similar to
Pattern A ($\dot{\mu}_\mathrm{int} > 0$), while the density of the
interface decreases ($\dot{\mu}_\mathrm{int} < 0$).  The difference
between pattern A and the pattern B growth is that in pattern A the core and
interface expand together relatively rapidly, while in pattern B the
core grows at the expense of the interface.

\item
Pattern C
\textit{(balanced growth).} As shown in
Fig.~\ref{fig:packing_group}c, the packing fraction of both the
interface and core are roughly constant through much of the folding in
Pattern C growth ($\dot{\mu}_\mathrm{int} \approx 0$ and
$\dot{\mu}_\mathrm{int} \approx 0$).  Here, as the nucleus grows, the
interfacial residues incorporated into the folded core are balanced by
unfolded residues entering the interfacial region.

\end{itemize}

The growth mode of the nucleus for the 27 proteins considered in this
paper (1pgb16 is too small to have a compact folded core) can be
roughly classified as follows: Pattern A (1pgb, 1a0n, 2ptl, 1shg, 1psf,
1pks, 1pin, 1c8c, 1fkb, 1fnf9, 1wit, 1urn); Pattern B
(2pdd, 1enh, 1coa, 1vii, 1aps, 1imq, 2abd, 1hdn,1 div); Pattern C (1lmb,
1csp, 1srl, 1ten, 1o6x, 1mef).

\section*{Polarized vs diffuse critical nucleus.}
\begin{figure*}[p]
\centering
\includegraphics[width=2.50in]{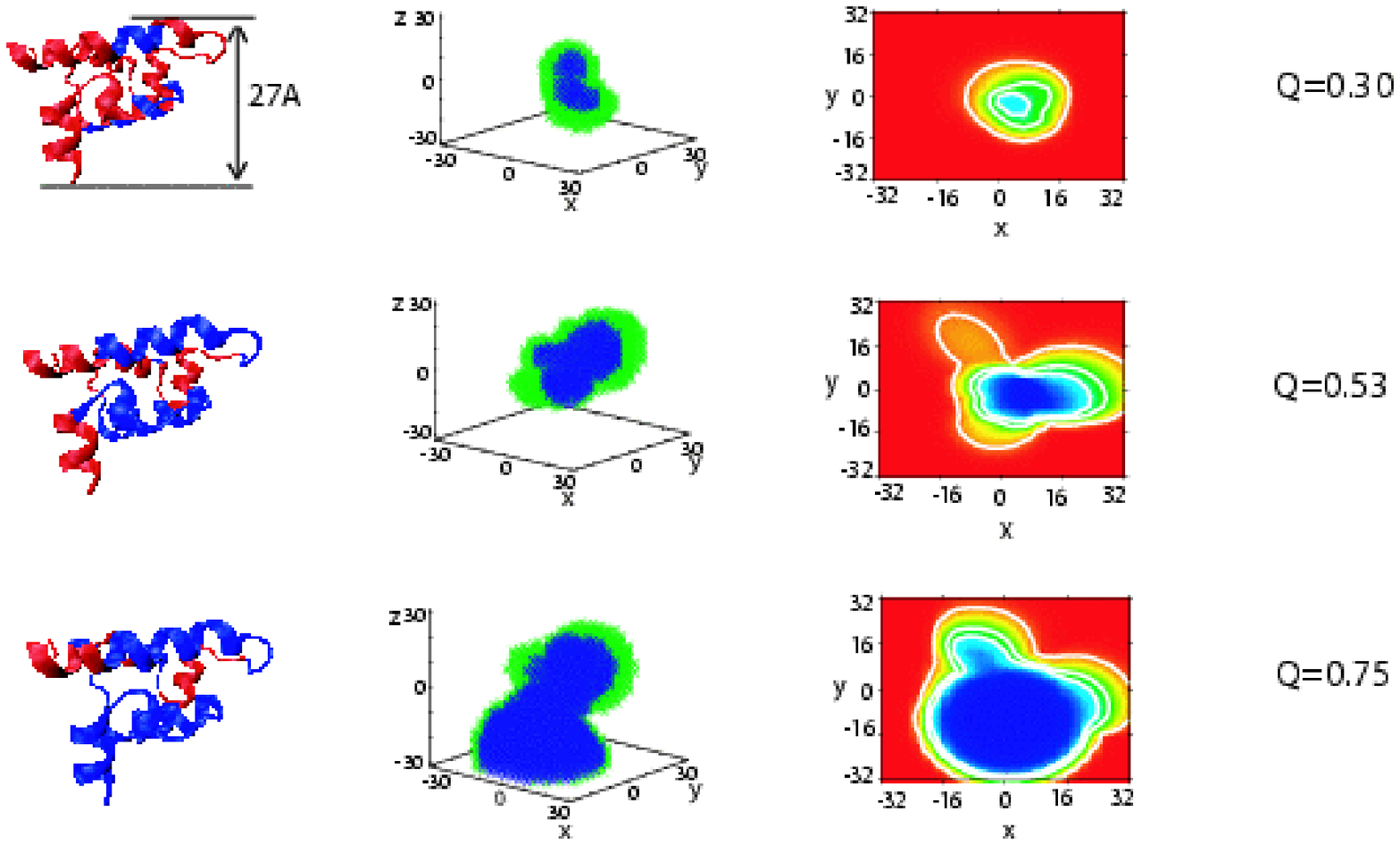}
\caption{Illustration of growth of folding nucleus and interface along the 
folding route (increasing $Q$) for the 
$\lambda$-repressor protein (1lmb).  Column 1 shows the 
three-dimensional folded structure with the fiducial set of 
folded residues colored blue and the unfolded residues colored red.
In Column 2, the folded core (colored blue) is surrounded by 
the interfacial region (colored green). Column 3 is a projection of
the indicator function $\tilde{n}(\mathbf{r})$ that defines the folded
and interfacial regions in space. The values correspond to 
$max_z \tilde{n}(x,y,z)$, ranging from 1 (blue) to 0 (red) in steps of 
0.01. Contour lines correspond to 0.1, 0.5, 0.7. Column 4 gives the 
corresponding Q value for each row. The critical nucleus
corresponds to $Q = 0.53$. The units for three plots are in Angstroms.
This protein is belong to the Pattern C (balanced growth). 
The three-dimensional structure was produced by VMD.~\cite{humphrey:schulten:96}
\label{fig:1lmb}}
\end{figure*}

\begin{figure}[p]
\centering
\includegraphics[width=2.50in]{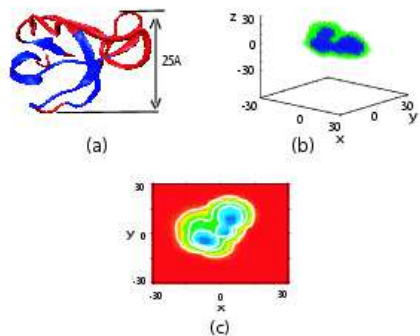}
\caption{An example of a polarized critical nucleus (Q=0.45) for Src-SH3 (1srl). 
Plots (a--c) correspond to the middle row of the diffuse nucleus shown in 
Fig.~\ref{fig:1lmb}.
\label{fig:1srl}}
\end{figure}

A folding mechanism is typically characterized by the structure of the critical
nucleus.  The spatial structure of the transition state ensemble,
inferred from $\phi$-value analysis, has often been qualitatively summarized
as either diffuse or polarized.~\cite{grantcharova:horwich:01}
Intermediate $\phi$-values spread across a large portion of the
protein sequence indicate a diffuse nucleus. In contrast, polarized
transition states are inferred when only one part of structure has
relatively high $\phi$-vales while the rest of the residues have low
$\phi$-values. In addition to a bimodal distribution of
$\phi$-values, the ordered residues in a polarized transition state
ensemble are located in one region in the native configuration.
Polarized and diffuse critical nuclei are sometimes called localized
and delocalized transition state ensembles,
respectively.~\cite{geierhaas:clarke:06} Of course, the critical
nucleus of a given protein is expected to have structural properties somewhere
between the two limits of polarized and diffuse.
The second row of Fig.~\ref{fig:1lmb} gives an example a diffuse
critical nucleus (1lmb).  For comparison, Fig.~\ref{fig:1srl} shows
the corresponding plots for a protein with a polarized critical
nucleus (1srl). Comparing with Fig.~\ref{fig:1lmb} and
Fig.~\ref{fig:1srl}, it is clear that the interface of 1lmb is much
broader than the interface region of 1srl. Furthermore, the folded
core of 1lmb is much more diffuse than the folded core of 1srl.

\begin{figure*}[p]
\centering
\includegraphics[width=2.50in]{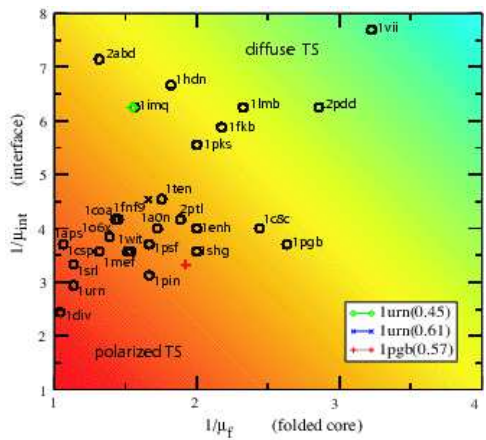}
\caption{Inverse packing fraction of the interface and folded core.
for 27 two-state proteins. 
The gradual change color shows the continuous
change from polarized nuclei (red) to diffuse nuclei (cyan).
Also shown are two early transition states of U1A (1urn) and
a late transition of protein G (1pgb).
\label{fig:sharpness}}
\end{figure*}

Characterizing a capillarity-like ordered nucleus as either diffuse or
polarized is a statement of the sharpness of interface as well as
compactness of the core.  For convenience, we monitor both regions by
the normalized volume per monomer (inverse packing fraction):
$1/\mu_\rmf$ and $1/\mu_\mathrm{int}$.  The results for the two state
proteins considered in this work are shown in
Fig.~\ref{fig:sharpness}.  Nuclei with small values of
$1/\mu_\mathrm{f}$ and $1/\mu_{\mathrm{int}}$ are more polarized with relatively compact
cores and sharp interfaces (similar to those envisioned in the strict capillarity
approximation).  Diffuse nuclei, on the other hand,
have extended regions of partial order which corresponds to large
values of $1/\mu_\mathrm{f}$ and/or $1/\mu_{\mathrm{int}}$. We note
that relatively polarized nuclei can
have cores that are still loosely packed compared to the native state
density (eg., 1pgb). Furthermore, relatively diffuse nuclei can have tightly
packed cores but extended interfaces (eg., 2abd, 1imq, 1fkb).


Our predictions for proteins with polarized transition state, such as
1csp~\cite{garcia-mira:schmid:04}, 1srl~\cite{riddle:baker:99},
1shg~\cite{martinez:serrano:99}, 1pin~\cite{jager:gruebele:01}
2ptl~\cite{kim:baker:00}, and 1pgb~\cite{mccallister:baker:00}, are
consistent with classification inferred by experimental $\phi-$value
analysis.  Several protein classified as having diffuse nuclei are
also predicted by our model, such as
1lmb~\cite{burton:oas:97},2abd~\cite{kragelund:poulsen:99},
1imq~\cite{friel:radford:03},and 1fkb~\cite{fulton:Jackson:99}.
Nevertheless, the predictions are at odds with experimental measurement
for a few proteins. Our model predicts that the folding nucleus of
CI2, perhaps the archetype for a diffuse transition state ensemble, is
relatively polarized.  This is also true for
1aps.~\cite{chiti:dobson:99} Another exception is U1A, which has been shown experimentally
to have an
early, delocalized transition state.~\cite{ternstrom:oliveberg:99} The calculated
folding route from this cooperative model
several transition states along the folding path, but the highest
barrier corresponds to a late, polarized nucleus.  If we look at the
earlier transition state ensembles (also shown in
Fig.~\ref{fig:sharpness}) the structure of the nucleus in much more
diffuse. The same situation arises in
1pgb,~\cite{mccallister:baker:00} for which the calculated folding
route has two transition states; the early one which is more diffuse
has a lower free energy than the late one which is more polarized.  

We note that in these exceptional cases, $\phi$-value distributions indicate
the critical nucleus is rather diffuse while our model predict more polarized nuclei.
This tendancy can be understood as a
consequence of the model being overly cooperative for these proteins,
since cooperativity generically tends to sharpen the interface between
folded region and unfolded region, and hence is somewhat biased
towards polarized transition states.


\section*{Conclusion}

In this paper, we directly characterize folding in terms of the
capillarity-like growth of the folding nucleus.  The nature of the
partially folded interfacial region between the folded core and
unfolded halo is the central focus of characterizing the growth modes
of the nucleus. We find that the growth of the nucleus can be classified into three
different patterns: (A) the core and interface both condense along the
folding route; (B) the core condenses at the expense of the
interfacial region; and (C) the growth of the core is balanced by
the monomers entering the interfacial region from the unfolded halo.
The picture of the core as close packing of rigid monomers appears to
be valid on average, though the size of the effective monomers is
larger than one would expect for a native-like, compact core. Furthermore, 
this analysis clarifies that diffuse nuclei inferred by the
distribution of intermediate $\phi$-values for example can arise from
either a diffuse folded core, a broad interfacial regions, or both.
The predictions from our calculations can be tested from the analysis of
the evolution of $\phi$-values as a function of the movement of the
transition state ensemble ($\beta^{\dagger}$) 
pioneered by Oliveberg and co-workers.~\cite{ternstrom:oliveberg:99,shen:wolynes:05}

The variational model considered here includes a uniform
``neutral'', excluded volume type cooperatively developed to account
of general trends in the absolute folding rates of two state
proteins.\cite{qi:portman:07} The exceptional qualitative
discrepancies of the the polarized versus diffuse characterization of
the critical nucleus (such as CI2, 1aps, 1pgb, and U1A) permit
an opportunity to assess the form and strength of the cooperativity
of this model.  The spatial density of the
critical nucleus can be used as an independent criteria to check the
value of the cooperativity obtained by simultaneously fitting
$\phi$-values and barrier height by the parameterization of the
cooperativity for each protein.  
There are some indications that one should consider
variations in the strength of the cooperativity for different proteins (though,
admittedly this is very closely tied to the specific form of the cooperativity in the model).
For example, Ejtehadi and Plotkin recently found
that the strength of cooperativity from three-body interactions can be
tuned for each protein to bring simulations of $\phi$-values into
better agreement with experimental
measurement.~\cite{ejtehadi:plotkin:04} Furthermore, detailed analysis from a
similar variational model suggests that the cooperativity of the U1A
protein is much lower than assumed in this
model.~\cite{shen:wolynes:05} The generally good qualitative agreement
between our calculations and experimental inferences about the spatial
extent of folding nuclei suggest that tuning the excluded volume
strength for each protein would not greatly improve the results
presented here for the majority of the proteins studied.

\begin{acknowledgments}
This work was supported in part by grant awarded by the Ohio Board of Regents
Research Challenge program.
\end{acknowledgments}




\end{document}